# Version 11

# Comorbidity and Anticomorbidity: Autocognitive developmental disorders of structured psychosocial stress

Rodrick Wallace, Ph.D.
The New York State Psychiatric Institute
*

December 12, 2003

## Abstract

We examine interacting cognitive modules of human biology which, in the asymptotic limit of long sequences of responses, define the output of an appropriate 'dual' information source. Applying a 'necessary condition' communication theory formalism roughly similar to that of Dretske, but focused entirely on long sequences of signals, we examine the regularities apparent in comorbid psychiatric and chronic physical disorders using an extension of recent perspectives on autoimmune disease. We find that structured psychosocial stress can literally write a distorted image of itself onto child development, resulting in a life course trajectory to characteristic forms of comorbid mind/body dysfunction affecting both dominant and subordinate populations within a pathogenic social hierarchy.

## I. Introduction

Certain mental disorders, for example depression and substance abuse, and many physical conditions like lupus, coronary heart disease, hypertension, breast and prostate cancers, diabetes, obesity, and asthma, show marked regularities at the community level of organization according to the social constructs of 'race', 'gender', 'ethnicity', and 'socioeconomic status'. Indeed, a virtual research industry has emerged in the United States to address the 'mystery' of such 'health disparities'. Population-level structure in disease permits profound insight into etiology because, to the extent these are 'environmental' disorders, the principal environment of humans is other humans, moderated by a uniquely characteristic embedding cultural context (e.g. Durham, 1991). Thus culturally-sculpted 'social exposures' are likely to be important at the individual, and critical at the population, levels of organization in the expression of certain mental disorders and a plethora of chronic diseases.

Further, mental disorders are often comorbidly expressed, both among themselves and with certain kinds of chronic physical disorder: Picture the obese, diabetic, depressed, anxious patient suffering from high blood pressure, asthma, coronary heart disease, and so on. Such comorbidity is the rule rather than the exception for the seriously ill, and is the central focus of this work. We will also examine how certain disorders may, in fact, be 'anticomorbid'.

Co- and antico- morbidity may indeed prove to be, for both medicine and public health, the kind of 'dirty open secret' that punctuation in the fossil record has proven to be for evolutionary theory (e.g. Gould, 2002; Eldredge, 1985). Just as open address of 'regular irregularities' in the fossil record has led to great advances in evolutionary thinking, more formal exploration of comorbidity may prove strikingly useful to contemporary medical and public health thinking.

As Cohen (2000) describes for autoimmune disease, however, the appearance of co- and antico- morbid conditions is, given the possibilities, rather surprisingly constrained to a relatively few often-recurring patterns. We will find this to be a central point.

Here we study how a long list of 'cognitive submodules' may become synergistically linked with embedding, culturally structured, psychosocial stress to produce comorbid patterns of illness associated with mental disorder and chronic disease. We will further suggest that many such disorders either have their roots in utero, as a stressed mother communicates environmental signals across the placenta, and programs her developing child's physiology, or else are initiated during early childhood. This pattern may affect underlying susceptibility to chronic infections or parasitic infestation as well as more 'systemic' disorders (e.g. Wallace and Wallace, 2002).

We are, then, particularly interested in the effects of 'stress' on the interaction between mind and body over the life course. 'Stress', we aver, is not often random in human societies, but is often itself a socially constructed cultural artifact, a very highly organized, 'language', having both a grammar and a syntax, so that certain stressors are 'meaningful' in a particular developmental context, and others are not, having little or no long-term physiological effect. We first argue that rational thought, emotion, immune function, related physiological processes like the hypothalamic-pituitary-adrenal (HPA) axis, blood pressure regulation, and sociocultural network function are, in fact, formally, if often weakly, cognitive systems. Each is associated with a 'dual information source' which may also be expressed as a kind of language. It is the punctuated in-terpenetration of these 'languages' which we will find critical to an understanding of how structured psychosocial stress affects the mind-body interaction, and, ultimately, writes a literal image of that structure upon that interaction, beginning in utero or early childhood, and determining a trajectory to inherently comorbid disease.

*Address correspondence to R. Wallace, PISCS Inc., 549 W. 123 St., Suite 16F, New York, NY, 10027. Telephone (212) 865-4766, rdwall@ix.netcom.com. Affiliation is for identification only.

We begin with a recitation of some formally cognitive submodules of human biology, in a large sense, which we believe interact both with each other and with structured psychosocial stress. Next we explore cognition as 'language', and infer the existence of a 'generalized cognitive homunculus' analogous to that explored by Cohen (2000) as the basis of autoimmune disease. Ultimately we propose a model based on autoimmune disease to account for a life trajectory of chronic comorbid psychiatric/physical disorder as involving a usually transient excited state of that homunculus which becomes a pathologically and recurrently 'permanent' zero-mode.

Although perhaps not falling strictly within our principal focus on the outcomes of structured psychosocial stress, schizophrenia seems amenable to an extension of our approach, most probably in a 'second order' manner, involving distortion of the spectrum of co- and antico- morbidities associated with the disorder. Fuller exploration of such a second order extension will be carried out in a subsequent work.

Some further comment on our methodology is appropriate.

We adapt recent advances in understanding 'punctuated equilibrium' in evolutionary process (e.g. Wallace, 2002b; Wallace and Wallace, 1998, 1999; Wallace et al., 2003) to the question of how embedding structured psychosocial stress affects the interaction of 'mind' and 'body', and specifically seek to determine how the synergism of such stress with cognitive submodule function might be constrained by certain of the asymptotic limit theorems of probability.

We know that, regardless of the probability distribution of a particular stochastic variate, the Central Limit Theorem ensures that long sums of independent realizations of that variate will follow a Normal distribution. Analogous constraints exist on the behavior of the 'information sources' we will find associated with both structured stress and cognitive module function – both independent and interacting – and these are described by the limit theorems of information theory. Imposition of phase transition formalism from statistical physics, in the spirit of the Large Deviations Program of applied probability, permits concise and unified description of evolutionary and cognitive 'learning plateaus' which, in the evolutionary case, are interpreted as evolutionary punctuation (e.g. Wallace, 2002a, b). This approach provides a 'natural' means of exploring punctuated processes in the effects of structured stress on mind-body interaction.

The model, as in the relation of the Central Limit Theorem to parametric statistical inference, is almost independent of the detailed structure of the interacting information sources inevitably associated with cognitive process, important as such structure may be in other contexts. This finesses some of the profound ambiguities associated with 'dynamic systems theory' and 'deterministic chaos' treatments in which the existence of 'dynamic attractors' depends on very specific kinds of differential equation models akin to those used to describe ecological population dynamics, chemical processes, or physical systems of weights-on-springs. Cognitive phenomena are neither well-stirred Erlenmeyer flasks of reacting agents, nor distorted mechanical clocks, and the application of 'nonlinear dynamic systems theory' to cognition will likely be found to involve little more than hopeful metaphor. Indeed, much of contemporary nonlinear dynamics can be subsumed within our formalism through 'symbolic dynamics' discretization techniques (e.g. Beck and Schlogl, 1995).

Rather than taking symbolic dynamics as an approximation to 'more exact' nonlinear ordinary or stochastic differential equation models, we throw out, as it were, the Cheshire cat and keep the cat's smile, generalizing symbolic dynamics to a more comprehensive information dynamics not constrained by 18th Century ghosts trapped in noisy, nonlinear, mechanical devices.

Our approach is conditioned, somewhat, by Waddington's (1972) vision that, in situations which arise when there is mutual interaction between the complexity-out-of-simplicity (self-assembly), and simplicity-out-of-complexity (self-organization), processes are to be discussed most profoundly with the help of the analogy of language, i.e. that *language* may become a paradigm go a Theory of General Biology, but a language in which basic sentences are programs, not statements.

In contrast to nonlinear systems theory approaches in which it appears impossible to actually write down the assumed underlying 'basic nonlinear equations' of cognitive phenomena, it does seem possible to uncover the grammar and syntax of both structured psychosocial stress and the function of cognitive submodules, and to express their relations in terms of empirically observed regression models relating measurable biomarkers, behaviors, beliefs, feelings, and so on.

Our analysis will focus on the eigenstructure of those models, constrained by the behavior of information sources under appropriate asymptotic limit theorems of probability.

Clearly, then, our approach takes much from parametric statistics, and, while idiosyncratic 'nonparametric' models may be required in special cases, we may well capture the essence of the most common relevant phenomena.

What we attempt is, in fact, surprisingly consonant with a current of mainstream thinking in cognitive science, what Adams (2003) characterizes as "the informational turn in philosophy", i.e. the relatively recent application of information-theoretic perspectives to the long, arduous, intellectual quest to understand 'mind'. One of the first reasonably successful syntheses was that of Dretske (1981, 1988, 1992, 1993, 1994), who put the matter thus (Dretske, 1994):

> "The mind can be viewed as an information-driven control system. To make this work, the idea of information must be operationalized in such a way as to give semantic properties (meaning, content) a role in the explanation of system behavior. This can be achieved by exploiting a statistical concept – mutual information – from communication theory. On this interpretation, some of the behavior of information-driven control systems is causally explained by the statistical correlations that exist between internal states and the external conditions about which they carry information...
>
> [A]lthough the chief concern of communication theory is the statistical properties of signals and channels, not the semantic information (if any) that these signals happen to carry over the channels, the statistical properties turn out to be relevant to what semantic information a signal can carry. Unless there is a statistically reliable channel of communication between [source] $S$ and [receiver] $R$, the signals reaching $R$ from $S$ cannot indicate what is happening at $S$... No signal can carry semantic information... unless the channel over which the signal arrives satisfies the appropriate statistical constraints of communication theory...

Communication theory can be interpreted as telling one something important about the conditions that are needed for the transmission of information as ordinarily understood, about what it takes for the transmission of semantic information. This has tempted people... to exploit [information theory] in semantic and cognitive studies, and thus, in the philosophy of mind...

If a semantic engine is a system whose performance is explained, not simply by the physical events occurring in it, but by the meaning of information that these events carry, then some systems, those capable of [learning], are semantic engines. The control system in such engines is an information-driven control system. Such systems are, in this sense and to this extent, minded"

That is, at very least, information theory explores the necessary conditions for cognitive process.

As Adams (2003) discusses in some detail, however, Dretske became particularly interested in questions regarding the information carried by a single symbol, i.e. the challenge of uniting the mathematical theory of information with a semantics:

"It is not uncommon to think that information is a commodity generated by things with minds. Let's say that a naturalized account puts matters the other way around, viz. it says that minds are things that come into being by purely natural causal means of exploiting the information in their environments. This is the approach taken by Dretske as he tried consciously to unite the cognitive sciences around the well-understood mathematical theory of communication...

[N]early everyone realized that information and mathematical properties of informational amounts and their transmission were not the same thing as semantic content or meaning... Dretske's insight was to see clearly (more clearly than most) what needed to be done to accomplish this."

Here we will finesse the problem by redirecting attention from the informational content or 'meaning' of individual symbols, i.e. the province of semantics, back to the statistical properties of long trains of symbols emitted by an 'information source'. As Dretske so clearly saw, this allows us to conduct scientific inference on the necessary conditions for cognitive process, but now in the asymptotic limit of very long sequences of output, which is quite precisely the home ground of classic communication theory. This redirection, we claim, will provide, along with some further modest asymptotic machinery imported from statistical physics, a sufficient basis for understanding the role of structured psychosocial stress in the etiology of developmental cognitive disorder, in a large sense.

### II. Some cognitive modules of human biology

**1. Immune function** Atlan and Cohen (1998) have proposed an information-theoretic cognitive model of immune function and process, a paradigm incorporating cognitive pattern recognition-and-response behaviors analogous to those of the central nervous system. This work follows in a very long tradition of speculation on the cognitive properties of the immune system (e.g. Tauber, 1998; Podolsky and Tauber, 1998; Grossman, 1989, 1992, 1993a, b, 2000).

From the Atlan/Cohen perspective, the meaning of an antigen can be reduced to the type of response the antigen generates. That is, the meaning of an antigen is functionally defined by the response of the immune system. The meaning of an antigen to the system is discernible in the type of immune response produced, not merely whether or not the antigen is perceived by the receptor repertoire. Because the meaning is defined by the type of response there is indeed a response repertoire and not only a receptor repertoire.

To account for immune interpretation Cohen (1992, 2000) has reformulated the cognitive paradigm for the immune system. The immune system can respond to a given antigen in various ways, it has 'options.' Thus the particular response we observe is the outcome of internal processes of weighing and integrating information about the antigen.

In contrast to Burnet's view of the immune response as a simple reflex, it is seen to exercise cognition by the interpolation of a level of information processing between the antigen stimulus and the immune response. A cognitive immune system organizes the information borne by the antigen stimulus within a given context and creates a format suitable for internal processing; the antigen and its context are transcribed internally into the 'chemical language' of the immune system.

The cognitive paradigm suggests a language metaphor to describe immune communication by a string of chemical signals. This metaphor is apt because the human and immune languages can be seen to manifest several similarities such as syntax and abstraction. Syntax, for example, enhances both linguistic and immune meaning.

Although individual words and even letters can have their own meanings, an unconnected subject or an unconnected predicate will tend to mean less than does the sentence generated by their connection.

The immune system creates a 'language' by linking two ontogenetically different classes of molecules in a syntactical fashion. One class of molecules are the T and B cell receptors for antigens. These molecules are not inherited, but are somatically generated in each individual. The other class of molecules responsible for internal information processing is encoded in the individual's germline.

Meaning, the chosen type of immune response, is the outcome of the concrete connection between the antigen subject and the germline predicate signals.

The transcription of the antigens into processed peptides embedded in a context of germline ancillary signals constitutes the functional 'language' of the immune system. Despite the logic of clonal selection, the immune system does not respond to antigens as they are, but to abstractions of antigens-in-context.

**2. Tumor control** We propose that the next cognitive submodule after the immune system is a tumor control mechanism which may include 'immune surveillance', but clearly transcends it. Nunney (1999) has explored cancer occurrence as a function of animal size, suggesting that in larger animals, whose lifespan grows as about the 4/10 power of their cell count, prevention of cancer in rapidly proliferating tissues becomes more difficult in proportion to size. Cancer control requires the development of additional mechanisms and systems to address tumorigenesis as body size increases – a synergistic effect of cell number and organism longevity. Nunney concludes

> "This pattern may represent a real barrier to the evolution of large, long-lived animals and predicts that those that do evolve ... have recruited additional controls [over those of smaller animals] to prevent cancer."

Different tissues may have evolved markedly different tumor control strategies. All of these, however, are likely to be energetically expensive, permeated with different complex signaling strategies, and subject to a multiplicity of reactions to signals, including those related to psychosocial stress. Forlenza and Baum (2000) explore the effects of stress on the full spectrum of tumor control, ranging from DNA damage and control, to apoptosis, immune surveillance, and mutation rate. Elsewhere (R. Wallace et al., 2003) we argue that this elaborate tumor control strategy, particularly in large animals, must be at least as cognitive as the immune system itself, which is one of its components: some comparison must be made with an internal picture of a 'healthy' cell, and a choice made as to response: none, attempt DNA repair, trigger programmed cell death, engage in full-blown immune attack. This is, from the Atlan/Cohen perspective, the essence of cognition.

**3. The HPA axis** The hypothalamic-pituitary-adrenal (HPA) axis, the 'flight-or-fight' system, is clearly cognitive in the Atlan/Cohen sense. Upon recognition of a new perturbation in the surrounding environment, memory and brain or emotional cognition evaluate and choose from several possible responses: no action needed, flight, fight, helplessness (i.e. flight or fight needed, but not possible). Upon appropriate conditioning, the HPA axis is able to accelerate the decision process, much as the immune system has a more efficient response to second pathogenic challenge once the initial infection has become encoded in immune memory. Certainly 'hyperreactivity' in the context of post-traumatic stress disorder (PTSD) is a well known example. Chronic HPA axis activation is deeply implicated in visceral obesity leading to diabetes and heart disease, via the leptin/cortisol diurnal cycle (e.g. Bjorntorp, 2001).

**4. Blood pressure regulation** Rau and Elbert (2001) review much of the literature on blood pressure regulation, particularly the interaction between baroreceptor activation and central nervous function. We paraphrase something of their analysis. The essential point, of course, is that unregulated blood pressure would be quickly fatal in any animal with a circulatory system, a matter as physiologically fundamental as tumor control. Much work over the years has elucidated some of the mechanisms involved: increase in arterial blood pressure stimulates the arterial baroreceptors which in turn elicit the baroreceptor reflex, causing a reduction in cardiac output and in peripheral resistance, returning pressure to its original level. The reflex, however, is not actually this simple: it may be inhibited through peripheral processes, for example under conditions of high metabolic demand. In addition, higher brain structures modulate this reflex arc, for instance when threat is detected and fight or flight responses are being prepared. This suggests, then, that blood pressure control cannot be a simple reflex, but is, rather, a broad and actively cognitive modular system which compares a set of incoming signals with an internal reference configuration, and then chooses an appropriate physiological level of blood pressure from a large repertory of possible levels, i.e. a cognitive process in the Atlan/Cohen sense. The baroreceptors and the baroreceptor reflex are, from this perspective, only one set of a complex array of components making up a larger and more comprehensive cognitive blood pressure regulatory module.

**5. Emotion** Thayer and Lane (2000) summarize the case for what can be described as a cognitive emotional process. Emotions, in their view, are an integrative index of individual adjustment to changing environmental demands, an organismal response to an environmental event that allows rapid mobilization of multiple subsystems. Emotions are the moment-to-moment output of a continuous sequence of behavior, organized around biologically important functions. These 'lawful' sequences have been termed 'behavioral systems' by Timberlake (1994).

Emotions are self-regulatory responses that allow the efficient coordination of the organism for goal-directed behavior. Specific emotions imply specific eliciting stimuli, specific action tendencies including selective attention to relevant stimuli, and specific reinforcers. When the system works properly, it allows for flexible adaptation of the organism to changing environmental demands, so that an emotional response represents a *selection* of an appropriate response and the inhibition of other less appropriate responses from a more or less broad behavioral repertoire of possible responses. Such 'choice', we will show, leads directly to something closely analogous to the Atlan and Cohen language metaphor.

Damasio (1998) concludes that emotion is the most complex expression of homeostatic regulatory systems. The results of emotion serve the purpose of survival even in nonminded organisms, operating along dimensions of approach or aversion, of appetition or withdrawal. Emotions protect the subject organism by avoiding predators or scaring them away, or by leading the organism to food and sex. Emotions often operate as a basic mechanism for making decisions without the labors of reason, that is, without resorting to deliberated considerations of facts, options, outcomes, and rules of logic. In humans learning can pair emotion with facts which describe the premises of a situation, the option taken relative to solving the problems inherent in a situation, and perhaps most importantly, the outcomes of choosing a certain option, both immediately and in the future. The pairing of emotion and fact remains in memory in such a way that when the facts are considered in deliberate reasoning when a similar situation is revisited, the paired emotion or some aspect of it can be reactivated. The recall, according to Damasio, allows emotion to exert its pairwise qualification effect, either as a conscious signal or as nonconscious bias, or both, In both types of action the emotions and the machinery underlying them play an important regulatory role in the life of the organism. This higher order role for emotion is still related to the needs of survival, albeit less apparently.

Thayer and Friedman (2002) argue, from a dynamic systems perspective, that failure of what they term 'inhibitory processes' which, among other things, direct emotional responses to environmental signals, is an important aspect of psychological and other disorder. Sensitization and inhibition, they claim, 'sculpt' the behavior of an organism to meet changing environmental demands. When these inhibitory processes are dysfunctional – choice fails – pathology appears at numerous levels of system function, from the cellular to the cognitive.

Thayer and Lane (2000) also take a dynamic systems perspective on emotion and behavioral subsystems which, in the service of goal-directed behavior and in the context of a behavioral system, they see organized into coordinated assemblages that can be described by a small number of control parameters, like the factors of factor analysis, revealing the latent structure among a set of questionnaire items thereby reducing or mapping the higher dimensional item space into a lower dimensional factor space. In their view, emotions may represent preferred configurations in a larger 'state-space' of a possible behavioral repertoire of the organism. From their perspective, disorders of affect represent a condition in which the individual is unable to select the appropriate response, or to inhibit the inappropriate response, so that the response selection mechanism is somehow corrupted.

Gilbert (2001) suggests that a canonical form of such 'corruption' is the excitation of modes that, in other circumstances, represent 'normal' evolutionary adaptations, a matter to which we will return at some length below.

**6. 'Rational thought'** Although a Cartesian dichotomy between 'rational thought' and 'emotion' may be increasingly suspect, nonetheless humans, like many other animals, do indeed conduct individual rational cognitive decision-making as most of us would commonly understand it. Various forms of dementia involve characteristic patterns of degradation in that ability.

**7. Sociocultural network** Humans are particularly noted for a hypersociality which inevitably enmeshes us all in group processes of decision, i.e. collective cognitive behavior within a social network, tinged by an embedding shared culture. For humans, culture is truly fundamental. Durham (1991) argues that genes and culture are two distinct but interacting systems of inheritance within human populations. Information of both kinds has influence, actual or potential, over behaviors, which creates a real and unambiguous symmetry between genes and phenotypes on the one hand, and culture and phenotypes, on the other. Genes and culture are best represented as two parallel lines or tracks of hereditary influence on phenotypes.

Much of hominid evolution can be characterized as an interweaving of genetic and cultural systems. Genes came to encode for increasing hypersociality, learning, and language skills. The most successful populations displayed increasingly complex structures that better aided in buffering the local environment (e.g. Bonner, 1980).

Successful human populations seem to have a core of tool usage, sophisticated language, oral tradition, mythology, music, and decision making skills focused on relatively small family/extended family social network groupings. More complex social structures are built on the periphery of this basic object (e.g. Richerson and Boyd, 1995). The human species' very identity may rest on its unique evolved capacities for social mediation and cultural transmission. These are particularly expressed through the cognitive decision making of small groups facing changing patterns of threat and opportunity, processes in which we are all embedded and all participate.

### III. Cognition as 'language'

Atlan and Cohen (1998) argue that the essence of cognition is comparison of a perceived external signal with an internal, learned picture of the world, and then, upon that comparison, the choice of one response from a much larger repertoire of possible responses. We make a very general model of this process.

Pattern recognition-and-response, as we characterize it, proceeds by convoluting an incoming external 'sensory' signal with an internal 'ongoing activity' – the 'learned picture of the world' – and, at some point, triggering an appropriate action based on a decision that the pattern of sensory activity requires a response. We need not specify how the pattern recognition system is 'trained', and hence we adopt a weak model, regardless of learning paradigm, which can itself be more formally described by the Rate Distortion Theorem. We will, fulfilling Atlan and Cohen's (1998) criterion of meaning-from-response, define a language's contextual meaning entirely in terms of system output.

The model, an extension of that presented in Wallace (2000), is as follows.

A pattern of 'sensory' input, say an ordered sequence $y_0, y_1, ...$, is mixed in a systematic way with internal 'ongoing' activity, the sequence $w_0, w_1, ...$, to create a path of composite signals $x = a_0, a_1, ..., a_n, ...$, where $a_j = f(y_j, w_j)$ for a function $f$. An explicit example will be given below. This path is then fed into a highly nonlinear 'decision oscillator' which generates an output $h(x)$ that is an element of one of two (presumably) disjoint sets $B_0$ and $B_1$. We take

$$B_0 \equiv b_0, ..., b_k,$$

$$B_1 \equiv b_{k+1}, ..., b_m.$$

Thus we permit a graded response, supposing that if

$$h(x) \in B_0$$

the pattern is not recognized, and if

$$h(x) \in B_1$$

the pattern is recognized and some action $b_j, k+1 \leq j \leq m$ takes place.

We are interested in composite paths $x$ which trigger pattern recognition-and-response exactly once. That is, given a fixed initial state $a_0$, such that $h(a_0) \in B_0$, we examine all possible subsequent paths $x$ beginning with $a_0$ and leading exactly once to the event $h(x) \in B_1$. Thus $h(a_0, ..., a_j) \in B_0$ for all $j < m$, but $h(a_0, ..., a_m) \in B_1$.

For each positive integer $n$ let $N(n)$ be the number of paths of length $n$ which begin with some particular $a_0$ having $h(a_0) \in B_0$ and lead to the condition $h(x) \in B_1$. We shall call such paths 'meaningful' and assume $N(n)$ to be considerably less than the number of all possible paths of length $n$ – pattern recognition-and-response is comparatively rare. We further assume that the finite limit

$$H \equiv \lim_{n \to \infty} \frac{\log[N(n)]}{n}$$

both exists and is independent of the path $x$. We will – not surprisingly – call such a cognitive process *ergodic*.

Note that disjoint partition of 'state space' may be possible according to sets of states which can be connected by 'meaningful' paths, leading to a 'natural' coset algebra of the system, a matter of some importance we will not pursue here.

We may thus define an ergodic information source **X** associated with stochastic variates $X_j$ having joint and conditional probabilities $P(a_0, ..., a_n)$ and $P(a_n|a_0, ..., a_{n-1})$ such that appropriate joint and conditional Shannon uncertainties may be defined which satisfy the relations (Cover and Thomas, 1991; Ash, 1990)

$$H[\mathbf{X}] = \lim_{n \to \infty} \frac{\log[N(n)]}{n} =$$

$$\lim_{n \to \infty} H(X_n|X_0, ..., X_{n-1}) =$$

$$\lim_{n \to \infty} \frac{H(X_0, ..., X_n)}{n}.$$

(1)

We say this information source is *dual* to the ergodic cognitive process.

The Shannon-McMillan Theorem provides a kind of 'law of large numbers' and permits definition of the Shannon uncertainties in terms of cross-sectional sums of the form

$$H = -\sum P_k \log[P_k],$$

where the $P_k$ are taken from a probability distribution, so that $\sum P_k = 1$. Again, Cover and Thomas (1991) or Ash (1990) provide algebraic details.

It is important to recognize that different 'languages' will be defined by different divisions of the total universe of possible responses into various pairs of sets $B_0$ and $B_1$, or by requiring more than one response in $B_1$ along a path. Like the use of different distortion measures in the Rate Distortion Theorem (e.g. Cover and Thomas, 1991), however, it seems obvious that the underlying dynamics will all be qualitatively similar. Nonetheless, dividing the full set of possible responses into the sets $B_0$ and $B_1$ may itself require 'higher order' cognitive decisions by another module or modules, suggesting the necessity of 'choice' within a more or less broad set of possible 'languages of thought'. This would directly reflect the need to 'shift gears' according to the different challenges faced by the organism, either cross-sectionally at a particular time, or developmentally as it matures, accounting for 'critical periods' in the onset of developmental disorder, a matter to which we will return. A critical problem then becomes the choice of a 'normal' zero-mode language among a very large set of possible languages representing the (hyper- or hypo-) 'excited states' accessible to the system. This is a fundamental point which we explore below in various ways.

In sum, meaningful paths – creating an inherent grammar and syntax – have been defined entirely in terms of system response, as Atlan and Cohen (1998) propose.

We can apply this formalism to the stochastic neuron in a neural network: A series of inputs $y_i^j, i = 1, ...m$ from $m$ nearby neurons at time $j$ to the neuron of interest is convoluted with 'weights' $w_i^j, i = 1, ..., m$, using an inner product

$$a_j = \mathbf{y}^j \cdot \mathbf{w}^j \equiv \sum_{i=1}^{m} y_i^j w_i^j$$

(2)

in the context of a 'transfer function' $f(\mathbf{y}^j \cdot \mathbf{w}^j)$ such that the probability of the neuron firing and having a discrete output $z^j = 1$ is $P(z^j = 1) = f(\mathbf{y}^j \cdot \mathbf{w}^j)$.

Thus the probability that the neuron does not fire at time j is just $1 - P$. In the usual terminology the $m$ values $y_i^j$ constitute the 'sensory activity' and the $m$ weights $w_i^j$ the 'ongoing activity' at time $j$, with $a_j = \mathbf{y}^j \cdot \mathbf{w}^j$ and the path $x \equiv a_0, a_1, ..., a_n, ...$. A more elaborate example is given in Wallace (2002a).

A little work leads to a standard neural network model in which the network is trained by appropriately varying **w** through least squares or other error minimization feedback. This can be shown to replicate rate distortion arguments, as we can use the error definition to define a distortion function which measures the difference between the training pattern $y$ and the network output $\hat{y}$ as a function, for example, of the inverse number of training cycles, $K$. As we will discuss in another context, 'learning plateau' behavior emerges naturally as a phase transition in the mutual information $I(Y, \hat{Y})$ driven by the parameter $K$.

Thus we will eventually parametrize the information source uncertainty of the dual information source to a cognitive pattern recognition-and-response with respect to one or more variates, writing, e.g. $H[\mathbf{K}]$, where $\mathbf{K} \equiv (K_1, ..., K_s)$ represents a vector in a parameter space. Let the vector **K** follow some path in time, i.e. trace out a generalized line or surface $\mathbf{K}(t)$. We will, following the argument of Wallace (2002b), assume that the probabilities defining $H$, for the most part, closely track changes in $\mathbf{K}(t)$, so that along a particular 'piece' of a path in parameter space the information source remains as close to memoryless and ergodic as is needed for the mathematics to work. Between pieces we impose phase transition characterized by a renormalization symmetry, in the sense of Wilson (1971). See Binney, et al. (1986) for a more complete discussion. Wallace et al. (2003) and R. Wallace and R.G. Wallace (2003) present detailed calculations of 'biological' renormalizations and 'universality class tuning' which take the theory well beyond simple physical analogs.

We will call such an information source 'adiabatically piecewise memoryless ergodic' (APME). The ergodic nature of the information sources is a generalization of the 'law of large numbers' and implies that the long-time averages we will need to calculate can, in fact, be closely approximated by averages across the probability spaces of those sources. This is no small matter.

Note that our treatment does not preclude the existence of cognitive processes or submodules which may not have appropriate dual information sources. We cannot, however, fit them easily into our development, although Wallace (2003) has begun to explore extension of the theory to a certain class of non-ergodic information sources.

## IV. Interacting information sources: 'sociocultural psychoneuroimmunology'

We suppose that the behavior of a cognitive subsystem can be represented by a sequence of 'states' in time, the 'path' $x \equiv x_0, x_1, ....$ Similarly, we assume an external signal of 'structured psychosocial stress' can also be represented by a path $y \equiv y_0, y_1, ....$ These paths are, however, both very highly structured and, within themselves, are serially correlated and can, in fact, be represented by 'information sources' $\mathbf{X}$ and $\mathbf{Y}$. We assume the cognitive process and external stressors interact, so that these sequences of states are not independent, but are jointly serially correlated. We can, then, define a path of sequential pairs as $z \equiv (x_0, y_0), (x_1, y_1), ....$

The essential content of the Joint Asymptotic Equipartition Theorem is that the set of joint paths $z$ can be partitioned into a relatively small set of high probability which is termed *jointly typical*, and a much larger set of vanishingly small probability. Further, according to the JAEPT, the *splitting criterion* between high and low probability sets of pairs is the mutual information

$$I(X,Y) = H(X) - H(X|Y) = H(X) + H(Y) - H(X,Y)$$

(3)

where $H(X), H(Y), H(X|Y)$ and $H(X,Y)$ are, respectively, the Shannon uncertainties of $X$ and $Y$, their conditional uncertainty, and their joint uncertainty. See Cover and Thomas (1991) or Ash (1990) for mathematical details. As stated above, the Shannon-McMillan Theorem and its variants permit expression of the various uncertainties in terms of cross sectional sums of terms of the form $-P_k \log[P_k]$ where the $P_k$ are appropriate direct or conditional probabilities. Similar approaches to neural process have been recently adopted by Dimitrov and Miller (2001).

The high probability pairs of paths are, in this formulation, all equiprobable, and if $N(n)$ is the number of jointly typical pairs of length $n$, then, according to the Shannon-McMillan Theorem and its 'joint' variants,

$$I(X,Y) = \lim_{n \to \infty} \frac{\log[N(n)]}{n}.$$

(4)

Generalizing the earlier language-on-a-network models of Wallace and Wallace (1998, 1999), we suppose there is a 'coupling parameter' $P$ representing the degree of linkage between the cognitive system of interest and the structured 'language' of external signals and stressors, and set $K = 1/P$, following the development of those earlier studies. Then we have

$$I[K] = \lim_{n \to \infty} \frac{\log[N(K,n)]}{n}.$$

The essential 'homology' between information theory and statistical mechanics lies in the similarity of this expression with the infinite volume limit of the free energy density. If $Z(K)$ is the statistical mechanics partition function derived from the system's Hamiltonian, then the free energy density is determined by the relation

$$F[K] = \lim_{V \to \infty} \frac{\log[Z(K)]}{V}.$$

(5)

$F$ is the free energy density, $V$ the system volume and $K = 1/T$, where $T$ is the system temperature.

We and others argue at some length (e.g. Wallace and Wallace, 1998, 1999; Wallace, 2000; Rojdestvensky and Cottam, 2000; Feynman, 1996) that this is indeed a systematic mathematical homology which, we contend, permits importation of renormalization symmetry into information theory. Imposition of invariance under renormalization on the mutual information splitting criterion $I(X,Y)$ implies the existence of phase transitions analogous to learning plateaus or punctuated evolutionary equilibria in the relations between cognitive mechanism and external perturbation. An extensive mathematical treatment of these ideas is presented elsewhere (Wallace, 2002b, Wallace et al., 2003).

Elaborate developments are possible. From a the more limited perspective of the Rate Distortion Theorem, a selective corollary of the Shannon-McMillan Theorem, we can view the onset of a punctuated interaction between the cognitive mechanism and external stressors as the literal writing of distorted image of those stressors upon cognition:

Suppose that two (piecewise, adiabatically memoryless) ergodic information sources **Y** and **B** begin to interact, to 'talk' to each other, i.e. to influence each other in some way so that it is possible, for example, to look at the output of **B** – strings $b$ – and infer something about the behavior of **Y** from it – strings $y$. We suppose it possible to define a retranslation from the B-language into the Y-language through a deterministic code book, and call $\hat{\mathbf{Y}}$ the translated information source, as mirrored by **B**.

Define some distortion measure comparing paths $y$ to paths $\hat{y}$, $d(y, \hat{y})$ (Cover and Thomas, 1991). We invoke the Rate Distortion Theorem's mutual information $I(Y, \hat{Y})$, which is the splitting criterion between high and low probability pairs of paths. Impose, now, a parametrization by an inverse coupling strength $K$, and a renormalization symmetry representing the global structure of the system coupling.

Extending the analyses, triplets of sequences, $Y_1, Y_2, Z$, for which one in particular, here $Z$, is the 'embedding context' affecting the other two, can also be divided by a splitting criterion into two sets, having high and low probabilities respectively. The probability of a particular triplet of sequences is then determined by the conditional probabilities

$$P(Y_1 = y^1, Y_2 = y^2, Z = z) = \Pi_{j=1}^n p(y_j^1|z_j)p(y_j^2|z_j)p(z_j).$$

(6)

That is, $Y_1$ and $Y_2$ are, in some measure, driven by their interaction with $Z$.

For large $n$ the number of triplet sequences in the high probability set will be determined by the relation (Cover and Thomas, 1992, p. 387)

$$N(n) \propto \exp[nI(Y_1; Y_2|Z)],$$

(7)

where splitting criterion is given by

$$I(Y_1; Y_2|Z) \equiv$$

$$H(Z) + H(Y_1|Z) + H(Y_2|Z) - H(Y_1, Y_2, Z).$$

We can then examine mixed cognitive/adaptive phase transitions analogous to learning plateaus (Wallace, 2002b) in the splitting criterion $I(Y_1, Y_2|Z)$. Note that our results are almost exactly parallel to the Eldredge/Gould model of evolutionary punctuated equilibrium (Eldredge, 1985; Gould, 2002).

We can, for the purposes of this work, extend this model to any number of interacting information sources, $Y_1, Y_2, ..., Y_s$ conditional on an external context $Z$ in terms of a splitting criterion defined by

$$I(Y_1; ...; Y_s|Z) = H(Z) + \sum_{j=1}^s H(Y_j|Z) - H(Y_1, ..., Y_s, Z),$$

(8)

where the conditional Shannon uncertainties $H(Y_j|Z)$ are determined by the appropriate direct and conditional probabilities.

### V. The simplest model of cognition and stress

Stress, as we envision it, is not a random sequence of perturbations, and is not independent of its perception. Rather, it involves a highly correlated, grammatical, syntactical process by which an embedding psychosocial environment communicates with an individual, particularly with that individual's multiple cognitive modules, typically in the context of social hierarchy. We first view the structured stress experienced by an individual as APME information source, interacting with a similar dual information source defined by some cognitive submodule or, more typically, a more complicated 'splitting criterion' defined by the interaction of several such modules.

Again, the ergodic nature of the 'language' of stress is essentially a generalization of the law of large numbers, so that long-time averages can be well approximated by cross-sectional expectations. Languages do not have simple autocorrelation patterns, in distinct contrast with the usual assumption of random perturbations by 'white noise' in the standard formulation of stochastic differential equations.

Let us suppose we cannot measure either stress or cognitive submodule function directly, but can determine the concentrations of hormones, neurotransmitters, certain cytokines, and other biomarkers, or else macroscopic behaviors, beliefs, feelings, and other responses associated with the function of cognitive submodules according to some 'natural' time frame inherent to the system. This would typically be the circadian cycle in both men and women, and the hormonal cycle in premenopausal women. Suppose, in the absence of extraordinary 'meaningful' psychosocial stress, we measure a series of $n$ biomarker concentrations, behavioral characteristics, and other indices at time $t$ which we represent as an $n$-dimensional vector $X_t$. Suppose we conduct a number of experiments, and create a regression model so that we can, in the absence of perturbation, write, to first order, the markers at time $t+1$ in terms of that at time $t$ using a matrix equation of the form

$$X_{t+1} \approx \mathbf{R}X_t + b_0,$$

(9)

where **R** is the matrix of regression coefficients and $b_0$ a (possibly zero) vector of constant terms.

We then suppose that, in the presence of a perturbation by structured stress

$$X_{t+1} = (\mathbf{R} + \delta\mathbf{R}_{t+1})X_t + b_0$$

$$\equiv \mathbf{R}X_t + \epsilon_{t+1},$$

(10)

where we have absorbed both $b_0$ and $\delta\mathbf{R}_{t+1}X_t$ into a vector $\epsilon_{t+1}$ of 'error' terms which are not necessarily small in this formulation. In addition it is important to realize that this is not a population process whose continuous analog is exponential growth. Rather what we examine is more akin to the passage of a signal – structured psychosocial stress – through a distorting physiological, psychological, or sociocultural filter.

If the matrix of regression coefficients **R** is sufficiently regular, we can (Jordan block) diagonalize it using the matrix of its column eigenvectors **Q**, writing

$$\mathbf{Q}X_{t+1} = (\mathbf{Q}\mathbf{R}\mathbf{Q}^{-1})\mathbf{Q}X_t + \mathbf{Q}\epsilon_{t+1},$$

(11)

or equivalently as

$$Y_{t+1} = \mathbf{J}Y_t + W_{t+1},$$

(12)

where $Y_t \equiv \mathbf{Q}X_t, W_{t+1} \equiv \mathbf{Q}\epsilon_{t+1}$, and $\mathbf{J} \equiv \mathbf{Q}\mathbf{R}\mathbf{Q}^{-1}$ is a (block) diagonal matrix in terms of the eigenvalues of **R**.

Thus the (rate distorted) writing of structured stress onto cognitive response through $\delta\mathbf{R}_{t+1}$ is reexpressed in terms of the vector $W_{t+1}$.

It is important to note that, in general, the eigenvectors of **R** are not orthogonal, suggesting the possibility that excitation of a single eigenvector will result in 'overtones' of 'mixed cognitive responses' in our model. We will return to the importance of nonorthogonality below.

The sequence of $W_{t+1}$ is the rate-distorted image of the information source defined by the system of external structured psychosocial stress. This formulation permits estimation of the long-term steady-state effects of that image on emotional state. The essential trick is to recognize that because everything is (APM) ergodic, we can either time or ensemble average both sides of equation (12), so that the one-period offset is absorbed in the averaging, giving an 'equilibrium' relation

$$<Y> = \mathbf{J}<Y> + <W>$$

or

$$<Y> = (\mathbf{I} - \mathbf{J})^{-1}<W>,$$

(13)

where **I** is the $n \times n$ identity matrix.

Now we reverse the argument: Suppose that $Y_k$ is chosen to be some fixed eigenvector of **R**. Using the diagonalization of **J** in terms of its eigenvalues, we obtain the average 'cognitive excitation' in terms of some eigentransformed pattern of exciting perturbations as

$$<Y_k> = \frac{1}{1-\lambda_k}<W_k>$$

(14)

where $\lambda_k$ is the eigenvalue of $<Y_k>$, and $<W_k>$ is some appropriately transformed set of ongoing perturbations by structured psychosocial stress.

The essence of this result is that *there will be a characteristic form of perturbation by structured psychosocial stress – the $W_k$ – which will resonantly excite a particular 'mixed cognitive eigenmode'*. Conversely, by 'tuning' the eigenmodes of **R**, output can be trained to galvanized response in the presence of particular forms of long-lasting perturbation.

This is because, if **R** has been appropriately determined from regression relations, then the $\lambda_k$ will be a kind of multiple correlation coefficient (e.g. Wallace and Wallace, 2000), so that particular eigenpatterns of perturbation will have greatly amplified impact. If $\lambda = 0$ then perturbation has no more effect than its own magnitude. If, however, $\lambda \to 1$, then the written image of a perturbing psychosocial stressor will have very great effect. Following Ives (1995), we call a system with $\lambda \approx 0$ *resilient* since its response is no greater than the perturbation itself.

In this model learning is, most obviously, the process of tuning response to perturbation. That is, we envision the regression matrix **R** as itself a tunable set of variables.

Suppose we require that $\lambda$ itself be a function of the magnitude of excitation, i.e.

$$\lambda = f(|<W>|)$$

where $|<W>|$ is the vector length of $<W>$. We can, for example, require the amplification factor $1/(1-\lambda)$ to have a signal transduction form, an inverted-U-shaped curve, for example the signal-to-noise ratio of a stochastic resonance, so that

$$\frac{1}{1-\lambda} = \frac{1/|<W>|^2}{1 + b\exp[1/(2|<W>|)]}.$$

(15)

This places particular constraints on the behavior of **R**, and gives a pattern of initial generalized hypersensitization, followed by anergy or 'burnout' with increasing average stress, a behavior that might well be characterized as 'pathological resilience', and may have multifactorial evolutionary significance: induced 'burnout' may become a viable game-theoretic or other strategy for particular forms of inescapable stress (e.g. Mealey, 1995).

### VI. The generalized cognitive homunculus and its retina: responding to sudden change

Cohen (2000) argues at some length for the existence of an 'immunological homunculus', i.e. the immune system's own perception of the body as a whole. The particular utility of such a homunculus, in his view, is that sensing perturbations in such a self-image can serve as an early warning sign of pending necessary inflammatory response – expressions of tumorigenesis, acute or chronic infection, parasitization, and the like. Thayer and Lane (2000) argue something analogous for emotional response as a quick internal index of larger patterns of threat or opportunity.

It seems obvious that the collection of interacting cognitive submodules we have explored above must also have a coherent internal self-image of the state of the mind-and-body and its social relationships. This inferred picture, at the individual level, we term the 'generalized cognitive homunculus', (GCH).

Suppose we write a GCH response to short-term perturbation – not the effects of long-lasting structured psychosocial stress – as

$$X_{t+1} = (\mathbf{R}_0 + \delta\mathbf{R}_{t+1})X_t.$$

Again we impose a (Jordan block) diagonalization in terms of the matrix of (generally nonorthogonal) eigenvectors $\mathbf{Q}_0$ of some 'zero reference state' $\mathbf{R}_0$, obtaining, for an initial condition which is an eigenvector $Y_t \equiv Y_k$ of $\mathbf{R}_0$,

$$Y_{t+1} = (\mathbf{J}_0 + \delta\mathbf{J}_{t+1})Y_k = \lambda_k Y_k + \delta Y_{t+1} =$$

$$\lambda_k Y_k + \sum_{j=1}^{n} a_j Y_j,$$

(16)

where $\mathbf{J}_0$ is a (block) diagonal matrix as above, $\delta\mathbf{J}_{t+1} \equiv \mathbf{Q}_0 \delta\mathbf{R}_{t+1}\mathbf{Q}_0^{-1}$, and $\delta Y_{t+1}$ *has been expanded in terms of a spectrum of the eigenvectors of* $\mathbf{R}_0$, with

$$|a_j| \ll |\lambda_k|, |a_{j+1}| \ll |a_j|.$$

(17)

The essential point is that, provided $\mathbf{R}_0$ has been properly 'tuned', so that this condition is true, the first few terms in the spectrum of the plieotropic iteration of the eigenstate will contain almost all of the essential information about the perturbation, i.e. most of the variance. We envision this as similar to the detection of color in the optical retina, where three overlapping non-orthogonal 'eigenmodes' of response suffice to characterize a vast array of color sensations. Here, if a concise spectral expansion is possible, a very small number of (typically nonorthogonal) 'generalized cognitive eigenmodes' permit characterization of a vast range of external perturbations, and rate distortion constraints become very manageable indeed. Thus GCH responses – the spectrum of excited eigenmodes of $\mathbf{R}_0$, provided it is properly tuned – can be a very accurate and precise gauge of environmental perturbation.

The choice of zero reference state $\mathbf{R}_0$, i.e. the 'base state' from which perturbations are measured, is, we claim, a highly nontrivial task, necessitating a specialized apparatus.

This is a critical point. According to current theory, the adapted human mind functions through the action and interaction of distinct mental modules which evolved fairly rapidly to help address special problems of environmental and social selection pressure faced by our Pleistocene ancestors (e.g. Barkow et al., 1992). Here we have postulated the necessity of other physiological and social cognitive modules. As is well known in computer engineering, calculation by specialized submodules – e.g. numeric processor chips – can be a far more efficient means of solving particular well-defined classes of problems than direct computation by a generalized system. We suggest, then, that our generalized cognition has evolved specialized submodules to speed the address of certain commonly recurring challenges.

We argue that identification of the 'normal' state of the GCH – generalized cognition's self-image of the body and its social relationships – is a difficult matter requiring a dedicated cognitive submodule within overall generalized cognition. This is essentially because, for the vast majority of information systems, unlike mechanical systems, there are no 'restoring springs' whose low energy state automatically identifies equilibrium: relatively speaking, all states of the GCH are 'high energy' states. That is, active comparison must be made of the state of the GCH with some stored internal reference picture, and a decision made about whether to reset to zero, which is a cognitive process. We further speculate that the complexity of such a submodule must also follow something like Nunney's power law with animal size, as the overall generalized cognition and its image of the self, become increasingly complicated with rising number of cells and levels of linked cognition.

Failure of that cognitive submodule results in identification of a usually transient activated state of the GCH as 'normal', triggering the collective patterns of systemic activation (possibly including persistent underactivation) which constitute certain comorbid mental and chronic physical disorders. This would result in a relatively small number of characteristic 'eigenforms' of comorbidity, which would typically become more mixed with increasing disorder.

In sum, since such 'zero mode identification' (ZMI) is a (presumed) cognitive submodule of overall generalized cognition, it involves convoluting incoming 'sensory' with 'ongoing' internal memory data in choosing the zero state, i.e. defining $\mathbf{R}_0$. The dual information source defined by this cognitive process can then interact in a punctuated manner with 'external information sources' according to the Rate Distortion and related arguments above. From a RDT perspective, then, those external information sources literally write a distorted image of themselves onto the ZMI, often in a punctuated manner: (relatively) sudden onset of a developmental trajectory to comorbid mental disorders and chronic physical disease.

Different systems of external signals – including but not limited to structured psychosocial stress – will, presumably, write different characteristic images of themselves onto the ZMI cognitive submodule, i.e. trigger different patterns of comorbid mental disorder and chronic diseases.

Elsewhere (R. Wallace, 2003) we speculate that patterns of autoimmune disease are likely to be related to both circadian and hormonal cycles, factors which may come into play in comorbidity of more general mental and chronic physical disorder.

Further theoretical development would introduce the 'generalized Onsager relation' analysis of gradient effects in driving parameters which affects system behavior between phase transitions (e.g. Wallace, 2002a). All these extensions remain to be done, and are not trivial.

### VII. Discussion and conclusions

**1. Generalized autocognitive disorder as a developmental disease** To reiterate, if $Y$ represents the information source dual to 'zero mode identification' in generalized cognition, and if $Z$ is the information source characterizing structured psychosocial stress, which serves as an embedding context, the 'mutual information' between them

$$I(Y;Z) = H(Y) - H(Y|Z) \tag{18}$$

serves as a splitting criterion for pairs of linked paths of states.

We suppose it possible to parametize the coupling between these interacting information sources by some 'inverse temperature', $K$, writing

$$I(Y;Z) = I[K], \tag{19}$$

with structured psychosocial stress as the embedding context.

Invocation of the mathematical homology between equations (4) and (5) permits imposition of renormalization formalism (Wallace, 2000; Wallace et al., 2003a) resulting in punctuated phase transition depending on $K$.

Socioculturally constructed and structured psychosocial stress, in this model having both 'grammar' and 'syntax', can be viewed as entraining the function of zero mode identification when the coupling with stress exceeds a threshold. More than one threshold appears likely, accounting in a sense for the often staged nature of 'environmentally caused' disorders. These should result in a series of collective, but highly systematic, 'tuning failures' which, in the Rate Distortion sense, represents a literal image of the structure of imposed psychosocial stress written upon the ability of the GCH to characterize a 'normal' mode of excitation, causing a mixed atypical and usually transient state to become permanent, producing comorbid mental and chronic physical disorder. As discussed above, this process may have both cross-sectional and longitudinal structure, with the latter accounting for 'critical periods' in the onset of developmental disorders.

Coronary heart disease (CHD) is already understood as a disease of development, which begins *in utero*. Work by Barker and colleagues, which we cited above, suggests that those who develop CHD grow differently from others, both in utero and during childhood. Slow growth during fetal life and infancy is followed by accelerated weight gain in childhood, setting a life history trajectory for CHD, type II diabetes, hypertension, and, of course, obesity. Barker (2002) concludes that slow fetal growth might also heighten the body's stress responses and increase vulnerability to poor living conditions later in life. Thus, in his view, CHD is a developmental disorder that originates through two widespread biological phenomena, developmental plasticity and compensatory growth, a conclusion consistent with the work of Smith et al. (1998), who found that deprivation in childhood influences risk of mortality from CHD in adulthood, although an additive influence of adult circumstances is seen in such cases.

Much of the CHD work particularly implicates certain kinds of hypertension as a developmental disorder. As Eriksson et al. (2000) put the matter,

> "The association between low birth weight and raised blood pressure in later life has now been reported in more than 50 published studies of men, women, and children. It has been shown to result from retarded fetal growth rather than premature birth. The 'fetal origins' hypothesis proposes that the association reflects permanent resetting of blood pressure by undernutrition in utero."

With regard to asthma, Wright et al. (1998) find prospective epidemiological studies showing that the newborn period is dominated by Th2 reactivity in response to allergens, and it is also evident that Th1 memory cells selectively develop shortly after birth, and persist into adulthood in non-atopic subjects. For most children who become allergic or asthmatic, the polarization of their immune systems into an atopic phenotype probably occurs during early childhood. There is evidence that parental reports of life stress are associated with subsequent onset of wheezing in children between birth and one year. It has been speculated that stress triggers hormones in the early months of life which may influence Th2 cell predominance, perhaps through a direct influence of stress hormones on the production of cytokines that are thought to modulate the direction of immune cell differentiation.

Work by Hirsch (2003) can be interpreted as suggesting that obesity, which is also seriously epidemic in the USA, is a developmental disorder with roots in utero or early childhood. Hirsch and others have developed a 'set point' or homeostatic theory of body weight, finding that it is the process which determines that 'set point' which needs examination, rather than the homeostasis itself, which is now fairly well understood. Hirsch concludes that the truly relevant question is not why obese people fail treatment, it is how their level of fat storage became elevated, a matter, he concludes, is probably rooted in infancy and childhood, when strong genetic determinants are shaping a still-plastic organism.

Somewhat less conclusively, a lively debate rages regarding various possible subforms of psychopathy, a mental disorder characterized by a long history manipulative, impulsive, and callous antisocial 'cheating' behavior. Mealy (1995) places the disorder in an evolutionary perspective as either a genetically determined or an acquired 'life history strategy' very similar to Nunney's (1999) analysis of cancer, albeit at the social rather than cellular level of interaction. Paris (1993) attempts to provide a comprehensive, integrative, biopsychosocial perturbed 'condition-development' model for personality disorders, while Lalumiere et al. (2001), by contrast, find evidence for a strict life-history strategy model, concluding, as a result of studies on children and adolescents, that "If psychopathy is a result of condition-development, the environmental triggers are likely to operate very early". The review by Herpertz et al. (2000) examines the hypothesis that pathologically neglectful parenting and early social rejection contribute to onset of the disorder, particularly in the context of 'individualistic' social structures (e.g. Cooke, 1996). We speculate that it is possible to place the 'social cheating' of psychopathy in the same context as Nunney's cellular cheating for cancer, consequently being subject to the standard pattern of gene-environment 'norms of reaction' which will emerge as structured psychosocial stress has impact over the course of child development, probably beginning *in utero*.

It almost goes without saying that the diagnosis of psychopathy (like other 'personality disorders') is very much concentrated in prison subpopulations, and these have marked ethnic and 'racial' structure.

Anxiety disorders have along history of attribution to developmental factors and early childhood exposures (e.g. Bandelow et al., 2002). More generally, Egle et al. (2002) find evidence that early biological and psychosocial stress in childhood is associated with long-term vulnerability to various mental and physical diseases. Research findings have, in their view, accumulated on those emotional, behavioral and psychobiological factors which are responsible for the mediation of lifelong consequences including increased risk of somatization and other mental disorders such as anxiety, depression and personality disorders. These often result in high-risk behaviors that are associated with physical disease – cardiovascular disorders, stroke, viral hepatitis, type 2 diabetes, chronic lung disease, as well as with aggressive behavior.

We are led to suggest that these case histories represent a far more general phenomenon in the etiology of the larger spectrum of chronic and comorbid mental and physical disorders, in the sense that structured psychosocial stress can literally write an image of itself upon the developing child, and if acute enough, on the adult, initiating trajectories to comorbid mental and chronic physical disorder.

**2. Schizophrenia: The spectrum of co- and anticomorbidities** Schizophrenia, although it does not seem to display as marked a 'health disparities' pattern as the other disorders which are the central focus of this work, appears nonetheless to fall broadly within the paradigm of a developmental cognitive disorder (e.g. Lewis and Levitt, 2002; Allin and Murray, 2002). Within the United Kingdom schizophrenia is, however, significantly more prevalent among Afro-Caribbean immigrants subject to chronic unemployment, early separation from parents, and perhaps racial discrimination, when compared with non-migrants of either majority or minority ethnicity (Mallett et al., 2002). For the U.S. there is some controversy as to the propensity of majority clinicians to overdiagnose schizophrenia among minority patients, perhaps masking underlying demographic patterns. As Gaughran et al. (2002) note, however, there is good evidence of immune activation in schizophrenia. Up to a third of patients has an autoimmune condition clinically unrelated to their psychiatric illness, and first degree relatives of people with schizophrenia also have increased incidence of autoimmune disease.

Torrey and Yolken (2001) note the similarities and contrasts between schizophrenia and rheumatoid arthritis. Both are chronic, persistent diseases displaying lifelong prevalence and a relapsing and remitting course. Both are felt to involve environmental insults occurring in genetically susceptible individuals, and their diagnosis depends upon syndromal diagnostic criteria which have been developed by committees and have changed over time. Many studies, however, have observed a striking inverse correlation – an 'anticomorbidity' – between the two diseases, although both are believed to run in families, with a population prevalence of about one percent. That is, people with schizophrenia seem less likely to suffer from rheumatoid arthritis, although perhaps more likely to suffer autoimmune disease in general.

This begins to resemble the 'eigenmode' patterns discussed above.

Grossman et al. (2003) describe how the recent emphasis on schizophrenia as a developmental disorder has focused on characterizing the role of non-genetic factors in the development of symptom patterns. Certain prenatal and perinatal environmental exposures, including maternal stress and malnourishment, and obstetric complications such as low birth weight, have been reported to be associated with increased susceptibility to the disorder. Increased incidence has also been reported in children born to mothers who experienced infection from influenza or rubella during the second trimester of pregnancy. Thus early neurodevelopmental processes may be compromised, laying groundwork for disorder when taxed by later developmental demands, for example those associated with the stressful periods of social development in childhood and adolescence.

Rothermundt et al. (2001) further summarize at some length the case for both the infection and autoimmune hypotheses regarding onset of schizophrenia.

Torrey and Yolken (2001) conclude that the negative association between schizophrenia and rheumatoid arthritis may depend on the timing of some critical exposure, e.g. that exposure in utero or childhood produces schizophrenia, while exposure in adulthood produces rheumatoid arthritis. A slightly different hypothesis, consistent with the mathematical exercises above, is that rheumatoid arthritis and schizophrenia characterize different atypical mixed eigenmodes falsely and recurrently identified as zero states by the progressive failure of the ZMI module discussed above. Such would tend to be mutually exclusive, although not absolutely so since the eigenmodes are not orthogonal.

A broadly similar pattern has been commented on by Karlsson et al. (2001), who found homologous sequences of the HERV-W family of endogenous retroviruses in the cerebrospinal fluid of newly-diagnosed individuals with schizophrenia and in other subjects having multiple sclerosis. Karlsson et al. (2001) speculate it is possible that individuals with schizophrenia and multiple sclerosis undergo the activation of similar retroviral sequences but differ in terms of genetically determined responses to the retroviral activation. Schizophrenia and multiple sclerosis are distinct clinical entities and have different pathological manifestations, gender ratios, and clinical courses, but share a number of epidemiological features including age of onset, seasons of birth, and geographic distributions. In addition, however, some patients display clinical manifestations of both diseases.

Similarly, rigorous studies by Dupont et al. (1986), Gulbinat et al. (1992) and Mortinsen (1989, 1994), which followed large Danish and Dutch cohorts of patients with schizophrenia, when adjusted for smoking patterns, showed marked and highly significant reduction in a broad variety of cancers. More recent work by Cohen et al. (2002) adjusted for age, race, gender, marital status, education, net worth, smoking, and hospitalization in the year before death, for a large US sample likewise found markedly reduced risk of cancer among persons diagnosed with schizophrenia. Catts and Catts (2000) speculate that such results are driven by hyperactivation of the p53 tumor suppressor/apoptosis gene during neurodevelopment, causing long-term developmental dysfunction, while Teunis et al. (2002) suggest, from animal model studies, that the hyperreactive dopaminergic system characteristic of schizophrenia inhibits tumor vascularization.

These examples, again, strongly suggest the 'nonorthogonal eigenmode' pattern of the mathematical model, in which the ZMI module, including both immune function and the larger system of tumor control mechanisms within a unified and broadly cognitive structure, fails in a systematic epigenetic manner, producing characteristic spectra of co- and antico-morbidities among different dysfunctions.

We are led to speculate that, at the population level, structured psychosocial stress will exert a 'higher order effect', producing different spectra of co- and antico- morbidities between schizophrenia and other disorders within powerful and marginalized subgroups. This prediction should be empirically testable.

**3. Zero mode identification as a general problem for 'languages of thought'** For those dubious of Generalized Cognitive Homunculus regression model arguments, a brief reformulation in terms of the abstract development of section III above may be of interest. Recall that the essential characteristic of cognition in our formalism involves a function $h$ which maps a (convolutional) path $x = a_0, a_1, ..., a_n, ...$ onto a member of one of two disjoint sets, $B_0$ or $B_1$. Thus respectively, either (1) $h(x) \in B_0$, implying no action taken, or (2), $h(x) \in B_1$, and some particular response is chosen from a large repertoire of possible responses. We discussed briefly the problem of defining these two disjoint sets, and suggested that some 'higher order cognitive module' might be needed to identify what constituted $B_0$, the set of 'normal' states. Again, this is because there is no low energy mode for information systems: virtually all states are more or less high energy states, and there is no way to identify a ground state using the physicist's favorite variational or other minimization arguments on energy.

Suppose that higher order cognitive module, which we now recognize as a kind of Zero Mode Identification, interacts with an embedding language of structured psychosocial stress (or other systemic perturbation) and, instantiating a Rate Distortion image of that embedding stress, begins to include one or more members of the set $B_1$ into the set $B_0$. Recurrent 'hits' on that aberrant state would be experienced as episodes of highly structured comorbid mind/body pathology.

Empirical tests of this hypothesis, however, quickly lead again into real-world regression models involving the interrelations of measurable biomarkers, beliefs, behaviors, reported feelings, and so on, requiring formalism much like that used in sections V and VI.

**4. Cautions and implications** Much of our reasoning has been based on a fairly elaborate mathematical model of cognitive process. Mathematical models of physiological, social, and other ecosystems – like those we present here – are notorious for their unreliability, instability, and oversimplification. As it is said, "all models are wrong, but some models are useful". The mathematical ecologist E.C. Pielou (1977, p. 106) finds the usefulness of models consists *not in answering questions but in raising them*, i.e. models can be used to inspire new field investigations and these are the only source of new knowledge as opposed to new speculation.

The speculations arising from our analysis are of some interest. In particular our speculation that a pattern of co- and antico- morbid mental and chronic physical disorder represents a pathological, ordinarily atypical or transient, state is consistent with theorizing in both autoimmune disease and mental disorder. Gilbert (2001), for example, uses an evolutionary approach to conclude that the relatively small number of evolved adaptive defense mechanisms, for example the flight-or-fight hypothalamic-pituitary-adrenal (HPA) axis, may become pathologically activated to produce mental disorder. He suggests that such evolved defenses, of which there is a limited number, can become pathological when they are too easily aroused or prolonged, are arrested (i.e. aroused but not expressed), or are ineffective. These might involve depression, anxious arousal, or heightened vigilance to threat, with the type of defense (e.g. flight, fight, submit, help seeking) being mirrored in particular symptom presentations.

Jones and Blackshaw (2000) likewise argue that behavioral similarities between humans and animals show that many psychiatric states are distortions of evolved behavior, a perspective which provides, in their view, a new etiological approach to psychiatry transcending current mainstream empirical and phenomenological approaches which are principally forms of symptom classification.

Although individual pathologies of both mind and body may predominate in particular cases, our work here attempts to encompass a broad spectrum of chronic diseases, emotional disorders, and classic cognitive dysfunction, in the context of the local sociocultural network so important in human biology, and to explore the particular effects of structured psychosocial stress in the development of comorbid mind/body symptom patterns over the life course.

To reiterate, comorbidity may well be to medicine what the 'dirty open secret' of punctuation in the fossil record is to evolutionary theory (e.g. Gould, 2002).

The model which emerged focuses on the 'eigenstructure' of a generalized cognitive homunculus, and particularly on failure of a higher cognitive module which permits identification of the 'zero mode' of such a homunculus. For certain classes of mind/body symptomatology, early experiences of exposure to structured psychosocial stress can trigger identification of a morbid, highly atypical, mode as the zero-reference state, and initiate a life course of co- or antico- morbid psychiatric and physical disorders. The most typical pattern we have in mind would involve individual and population-level comorbidity among obesity, asthma, diabetes, hypertension, depression, anxiety, substance abuse, ruthless or violent behaviors, coronary heart disease, certain cancers, and asthma or lupus – what might well be characterized as 'oppression disorder' at the individual level.

Our analysis suggests that historical patterns of discrimination, deprivation, and injustice – for example the evolved system of slavery which has been characterized as 'American Apartheid' – are a determining feature in population-level expression of comorbid psychiatric and chronic physical disorder, patterns which are literally an image of that system imposed upon children, beginning in their mothers' wombs.

The more dubious aspects of the history of the United States are alive and well and being rewritten daily upon the developing bodies of its children.

This being said, the works of Albert Memmi (e.g. 1969) and Franz Fanon (e.g. 1966) show clearly that, for Apartheid systems, the reflective nature of structured psychosocial stress ensures that the health and welfare of both dominant and dominated populations will be closely linked through a wide variety of mechanisms. Within the United States this dynamic is clearly demonstrated by the relation between the obesity 'epidemics' within 'Black' and 'White' populations. Figure 1 shows the national Black death rate from diabetes (per 100,000) as a function of the White rate over the period 1979 to 1997. Both are rising rapidly, and, while that for Blacks is typically 1.5 times that for Whites, the increases are nonetheless very highly correlated indeed: $R^2 = 0.99$.

The 'obesity epidemic' in the US, then, appears to be a single system which enmeshes both populations, suggesting more generally that the health of those at the bottom of a draconian social hierarchy is often a leading indicator for the health of all, and that the health of those at the top appears seriously limited by the very structures which impose or exacerbate hierarchy. In this regard, figure 2 shows the rapidly increasing percent of total US income accounted for by the highest five percent of the population as a function of the integral of the number of manufacturing jobs lost in the US since 1980. The latter is an index of permanently dispersed social, economic, and political capital for large sectors of the population. The inference, of course, is that deindustrialization and its associated phenomena have exacerbated a pathogenic social hierarchy enmeshing both dominant and subordinate populations into trajectories of developmental disorder. Absent large-scale social and economic reform, this process will certainly continue.

## Acknowledgments

This work benefited from support under NIEHS Grant I-P50-ES09600-05 and from earlier funds provided through an Investigator Award in Health Policy Research from the Robert Wood Johnson Foundation. The author thanks Drs. D.N. Wallace, E. Struening, and J. Hirsch for useful discussions.

## Figure Captions

**Figure 1** US Black vs. White diabetes death rates (per 100,000), 1979-1997. While the Black rates are uniformly higher than the White, the coupling between them is very strong indeed, suggesting that, in the words of one researcher, "concentration is not containment" for chronic as well as for infectious diseases like AIDS or tuberculosis.

**Figure 2** Percent of total US income accounted for by the highest five percent of the population vs. integral of manufacturing job loss since 1980, 1980-1998.

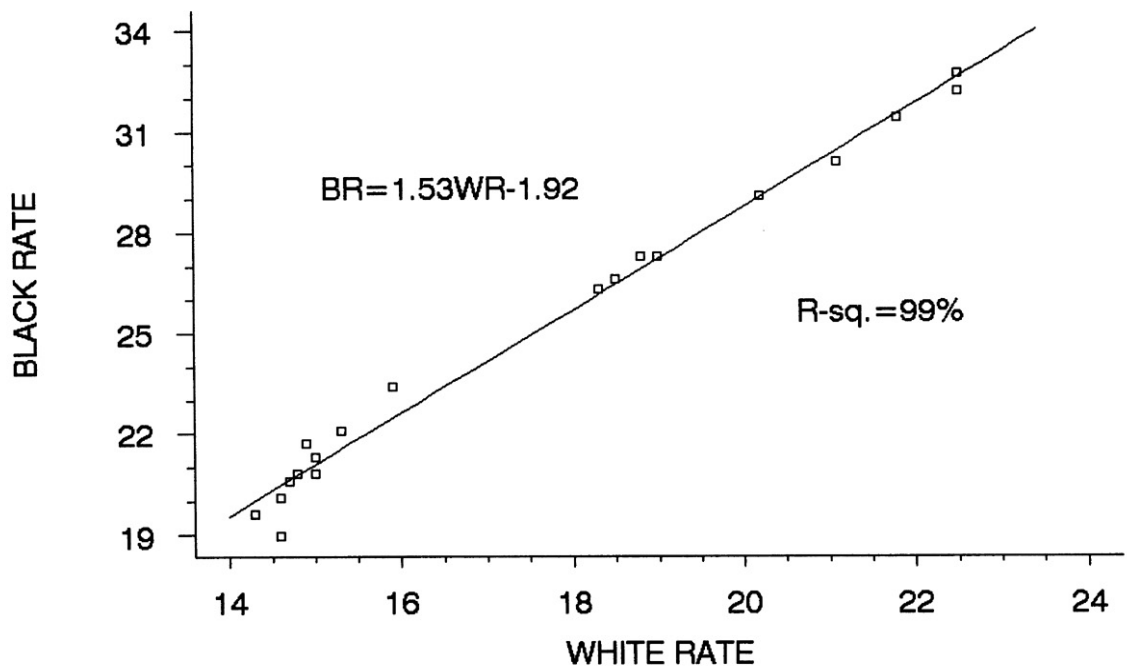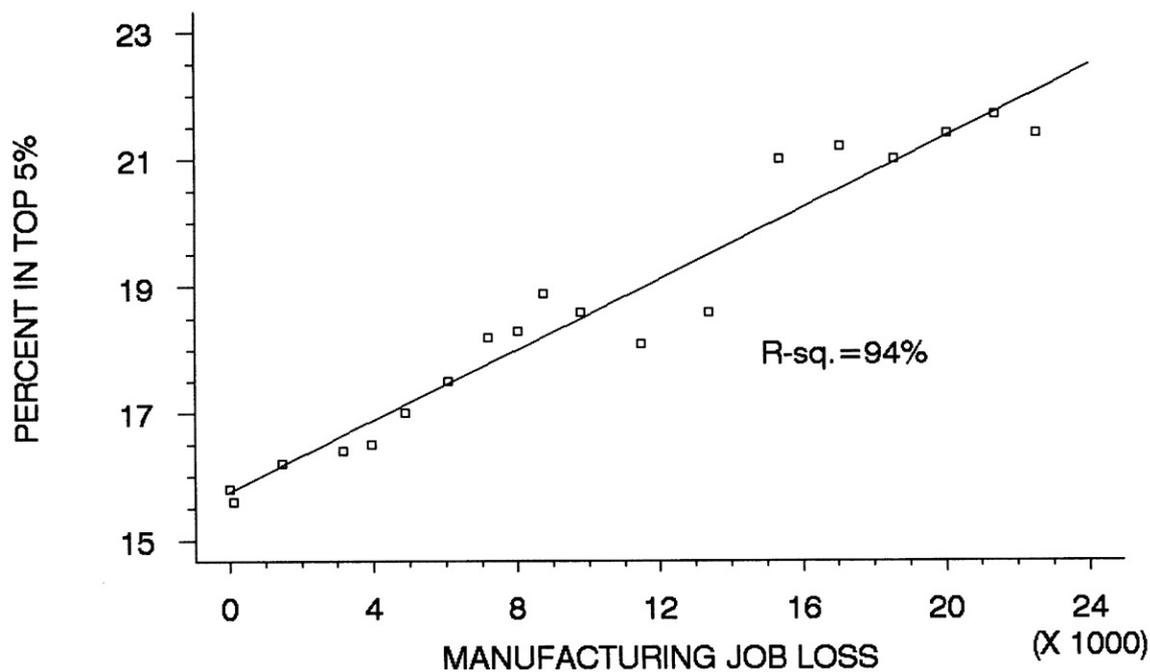